\documentclass{elsart5}

 \usepackage{graphicx}

\usepackage{amssymb}

\begin{document}

\begin{frontmatter}
% Title, authors and addresses
% use the thanksref command within \title, \author or \address for footnotes;
% use the corauthref command within \author for corresponding author footnotes;
% use the ead command for the email address,
% and the form \ead[url] for the home page:

\title{Energy of bond defects in quantum spin chains obtained from local approximations and from exact diagonalization}

%\title{Title\thanksref{tit1}}
%\thanks[tit1]{Title footnote}

%\author[aff1]{V. L. L\'{\i}bero\corauthref{cor1}\thanksref{label1}\ead{}}
\author{V. L. L\'{\i}bero\corauthref{cor1}}
\ead{libero@if.sc.usp.br}
%\ead[url]{home page}
%\thanks[label1]{author footnote}
\corauth[cor1]{Tel+55-16-33739854; fax+55-16-33739877}
\author{K. Capelle}
\author{F. C. Souza}
\author{A. P. F\'avaro}
\address{Instituto de F\'{\i}sica de S\~ao Carlos, Universidade de S\~ao
Paulo, 13560-970 S\~ao Carlos, SP, Brazil}
%\address[aff2]{Address}
%\thanks[label2]{aff footnote}
%\received{12 June 2005}
%\revised{13 June 2005}
%\accepted{14 June 2005}
%use optional labels to link authors explicitly to addresses:

%\author{}
%\address{}

\begin{abstract}
We study the influence of ferromagnetic and antiferromagnetic bond defects on 
the ground-state energy of antiferromagnetic spin chains. In the absence of 
translational invariance, the energy spectrum of the full Hamiltonian is
 obtained numerically, by an iterative modification of the power algorithm. 
In parallel, approximate analytical energies are obtained from a local-bond 
approximation, proposed here. This approximation results in significant 
improvement upon the mean-field approximation, at negligible extra 
computational effort.
\end{abstract}

%%%%%%%%%use  the \KEY command at the begin of keyword text%%%%%%%%%
\end{frontmatter}

\section{Introduction}\label{intro}

Many different types of magnetic systems are commonly modeled by spin 
Hamiltonians, such as the Heisenberg model. In special limits, such model 
Hamiltonians can be solved analytically, e.g., by the Bethe Ansatz for 
integrable one-dimensional systems. Away from special limits, powerful 
numerical methods, such as Monte Carlo simulations or density-matrix 
renormalization group techniques, provide much useful information, but at 
high computational cost. Full exact diagonalization is even more expensive, 
and thus limited to rather small systems. Quite generally, analytical and 
numerical techniques work best for models in which all sites are equivalent, 
so that translational invariance can be used to reduce the complexity of the 
problem. 

In magnetic nanostructures\cite{1} and molecular magnets\cite{2}, the generic 
many-body features of spin Hamiltonians coexist with additional real-life 
complications, such as impurities, defects, boundaries, textures, etc. The 
resulting {\it spatially inhomogeneous} spin Hamiltonians do not have 
translational symmetry, and often also lack other simplifying symmetries, 
which makes them hard to treat by traditional analytical and numerical methods.

Mean-field theory can, of course, always be applied, but is not sufficiently 
reliable to permit quantitative modeling, and is often even qualitatively 
wrong. In this paper we describe one analytical and one numerical method to 
obtain beyond-mean-field energies for  spin chains without translational 
invariance. 

The analytical calculations are based on model density-functional theory 
(DFT). In {\it ab initio} electronic-structure calculations DFT\cite{3} is a 
useful way to include correlations beyond the mean-field approximations, at very little 
additional computational cost. To deal with defects within DFT for spin
chains, we propose a local-bond approximation (LBA), akin to the local-density
approximation (LDA) of {\it ab initio} DFT and the local-spin approximation 
(LSA) previously proposed for spin systems with impurities\cite{4,5,6,7}. 
These schemes are described in Sec. 2.

To obtain results with higher precision than is possible by local approximations, and to judge the performance of such simple improvements on mean-field theory, we also perform fully numerical calculations, employing an iterative modification of the power algorithm, described in Section 3. Representative results are discussed in Section 4.  

\section{Local approximations for spin Hamiltonians}

DFT has been applied to model Hamiltonians, such as the 
Heisenberg\cite{4,5,6,7} and the Hubbard\cite{8} model, within generalized 
local-density approximations. 
Specifically for the Heisenberg model with impurity spins, this scheme 
consists in adding to the mean-field energy a local approximation to the 
correlation energy $E_c$, of the form
\begin{equation}
E_c^{LSA}[J,S_i] = \sum_i e_c^{hom}(J,S) \mid_{S\rightarrow S_i}
\end{equation} 
where the sum runs over all sites $i$, and $e_c^{hom}(J,S)$ is the per-site 
correlation energy of the homogeneous spin $S$ Heisenberg model, for which 
various approximations exist\cite{4,5}. This so-called local-spin 
approximation\cite{4,5} has been applied to investigate the energetics of impurities in one, 
two and three-dimensional Heisenberg models\cite{4,5,6,7}. Impurities here are 
defined, relative to a homogeneous model in which all sites have the same 
spin $S$, as sites with a spin $S_I \ne S$ (see inset of Fig. 1 for an 
example).

Another important class of spatial inhomogeneity, defects, cannot be treated 
within the LSA. We define defects, again with respect to a homogeneous model, 
as bonds having strength  $J_D\ne J$ (see insets of Figs. 1 and  2 for 
examples). To extend the local-approximation scheme to spin Hamiltonians with defects, we here propose the {\it local-bond approximation (LBA)},
\begin{equation}
E_c^{LBA}[J_b,S] = \sum_b e_c^{hom}(J,S)|_{J\rightarrow J_b} \;,
\end{equation}
where the sum runs over all bonds $b$, and for $e_c^{hom}(J,S)$ we use the 
same expressions 
employed in the LSA. Conceptually, the LBA decomposes the system in bonds, 
acting between sites, whereas the LSA decomposes it in sites, connected by 
bonds. Both approximation schemes become exact for infinite homogeneous 
systems, and both can be used in analytical calculations, as all one has to 
do to obtain corrections to the mean-field energies is to evaluate Eqs. (1) 
or (2) site by site or bond by bond. 

\section{Numerical Ground-State Energy}

    To obtain energies of higher quality than is possible with analytical 
calculations employing local approximations, we resort to a numerical scheme. 
Even in the presence of impurity spins and/or bond defects, the Heisenberg 
Hamiltonian 
\begin{equation}
\hat{H} = \sum_i J_i \hat{S}_i \cdot \hat{S}_{i+1} \;,
\end{equation}
where $J_i$ is the exchange integral between nearest-neighbor spin-vectors 
$\hat{S}_i$ and $\hat{S}_{i+1}$, conserves the $z$-component 
of the total spin, $\hat{S}_z$. The Lieb-Mattis theorem guaranties that the GS
pertains to the subspace of minimum $|S_z|$. As basis vectors we therefore use 
the set of quantum numbers $\{|m_1,m_2,...,m_N \rangle \}$, where  $m_i$ are 
the eigenvalues of $\hat{S}_i$, and can take values $-S_i, -S_i+1,...,S_i$. A 
practical way of generating this set of states is by decomposing integer 
numbers. For instance, for a chain with four spins 1/2, the binary 
decomposition of the integer 6 gives the sequence 0110, representing the 
vector $|-1/2,1/2,1/2,-1/2\rangle$ , one among other five of the subspace with
$S_z=0$. It is straightforward to assemble the matrix representation of
$\hat{H}$ in this base. However, although the resulting matrix is sparse, 
storing it is impractical: for 20 spins 1/2 the order of the matrix is 184756,
with 1108536 non-vanishing elements. In practice, we therefore do not store 
the matrix, but compute each matrix element every time it is needed; this 
saves a lot of memory, but is time consuming. 

To obtain the ground-state energy of $\hat{H}$ we propose an iterative 
modification of the well-known power algorithm. This modification, to be 
described in more detail in a separate publication, is easy to code and 
requires less memory than the Lanczos method, although it typically takes 
more processing time to extract the ground state. The power algorithm starts 
by decomposing a trial function for the ground-state eigenvector in terms of 
the unknown eigenvectors $|\psi_n\rangle$ of $\hat{H}$, according to
$|\psi_T\rangle = \sum_n \alpha_n |\psi_n\rangle$, with $\alpha_n$ constants. 
If $\epsilon$ is an upper limit of the energy spectrum, acting $k$ times with 
the operator $\hat{H}-\epsilon$ on $|\psi_T\rangle$ yields 
\begin{equation}
(\hat{H}-\epsilon)^k |\psi_T\rangle = 
\sum_n \alpha_n (E_n-\epsilon)^k |\psi_n\rangle \;,
\end{equation}
where $\hat{H}|\psi_n\rangle = E_n |\psi_n\rangle$. For antiferromagnetic 
chains, the highest energy $\epsilon$ corresponds to the ferromagnetic 
configuration, whose value is trivial even in the presence of impurities or 
defects. For $k\rightarrow \infty$, the above series is dominated by the 
ground-state term $\alpha_0 (E_0-\epsilon)^k |\psi_0\rangle$. Therefore, we 
can extract $E_0$ by performing
\begin{equation}
E_0 = \lim_{k\rightarrow \infty} 
\frac{\langle \phi|(\hat{H}-\epsilon)^{k+1}|\psi_T\rangle}
{\langle \phi|(\hat{H}-\epsilon)^k|\psi_T \rangle} + \epsilon \;,
\end{equation}
where $|\phi\rangle$ is any vector non-orthogonal to the ground state. Without
the constant $\epsilon$, the method yields the highest (in modulus) eigenvalue
instead. Our experience shows that the best trial function is the N\'eel 
state, a mean-field approximation for the ground state, consisting of a 
sequence of an up and down spins.

   To speed up the search for $E_0$, we implement the above limit iteratively:
at each step $k$ , we use for $|\phi\rangle$ the state $|\phi_{k-1}\rangle=
(\hat{H}-\epsilon)^{k-1}|\psi_T\rangle$ obtained in the previous step. This 
reduces the processing time to reach convergence, which we characterize by two
successive values of $E_0/N$ differing by less than $10^{-13}$. Typically, a 
few hundred $k$-steps are required for chains larger than 20 spins, and only a
few dozens for smaller chains. Using a desktop microcomputer with 1.5 MB of 
RAM, we obtained the ground-state energy of a homogeneous chain with 30 spins 
1/2, reproducing the results of Ref.\cite{9}. Bond defects do not require 
more memory space. 

\section{Antiferromagnetic spin chains with ferro- and antiferromagnetic 
defects}

As a first application of the LBA concept we compare, in Fig. 1, an 
antiferromagnetic (AFM) spin 1/2  ring (periodic boundary conditions) with 
one impurity spin $S_I=3/2$  to an AFM spin 1/2 ring with one bond defect
$J_D=5J$. In the mean-field approximation, both systems are, erroneously, 
predicted to have the same ground-state energy. This spurious degeneracy is 
lifted by adding the LSA and LBA correlation energies, respectively. Judged 
by the remaining distance to the exact data, LSA performs slightly better for 
the impurity than LBA does for the defect, but both provide significant 
improvements on the mean-field data.

\begin{figure}[h]
\centering
\includegraphics[scale=0.9]{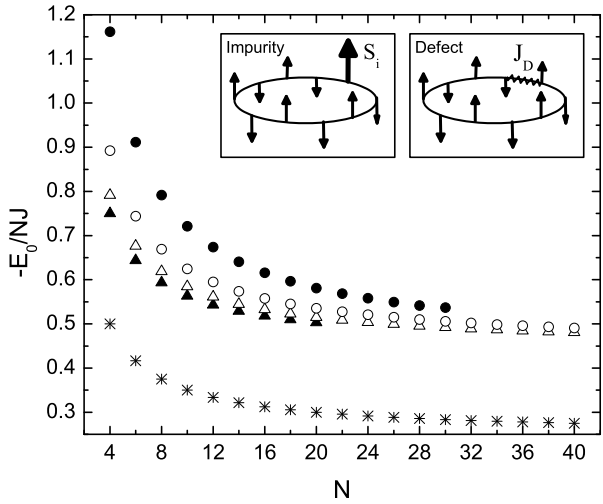}
\caption{Ground-state energy of an AFM spin 1/2 ring  
with a spin 3/2 impurity (triangles), and with an AFM defect of strength  
$J_D=5J$ (circles), treated exactly  (filled symbols) and via local 
approximations LSA/LBA (open symbols). Within mean-field theory (stars), the 
impurity and  the defect yield the same ground-state energy. Inclusion of 
correlation energy in the LSA (impurity case, open triangles) and the LBA 
(defect case, open circles) removes this spurious degeneracy.}
\end{figure}

Ferromagnetic (FM) defects, with $J_D < 0$, can be handled similarly. However,
an ambiguity arises in how the local substitution is to be performed. In 
principle, the LBA can be applied to a FM defect by substituting 
$J \rightarrow |J_i|$, $J \rightarrow J_i$ or 
$J \rightarrow \delta_{J_i,|J_i|}$. The first possibility can lead to positive
correlation energies, in violation of the variational principle. The second 
and third possibilities correctly predict negative correlation energies, but 
numerically the second is slightly inferior to the third for the type of 
system investigated here, as judged by comparison to exact data. Below, we 
thus employ the third substitution. Physically, this choice,
$J \rightarrow \delta_{J_i,|J_i|}$, means that a ferromagnetic bond does not 
contribute to the correlation energy, which is rather reasonable, as at zero 
temperature the mean-field energy of a homogeneous ferromagnetic chain is 
already exact.

In Fig. 2 we show ground-state energies of an AFM spin chain (open boundary 
conditions) with one AFM defect $J_D=+3J$, and of the same chain with one 
relaxed FM 
defect, $J_D=-3J$. Comparison of LBA data with exact data shows that the LBA 
significantly improves on the mean-field approximation both for FM and AFM 
defects. Quantitatively, the performance for the FM defect (squares in Fig. 2)
is better than that for the AFM defect, but in both cases a significant
improvement over the mean-field data is achieved. Note, in particular, that 
for an FM defect the mean-field curve is even qualitatively wrong, predicting 
a wrong sign for the slope at $N\rightarrow 0$, whereas the LBA recovers the 
correct behavior.

\begin{figure}[h]
\centering
\includegraphics[scale=0.9]{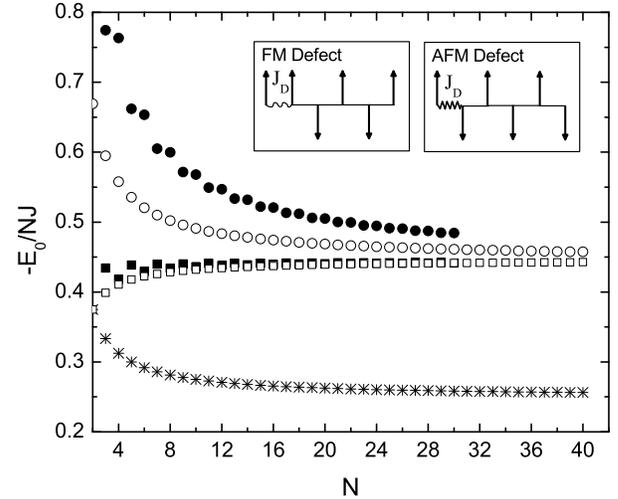}
\caption{Ground-state energy of an AFM spin 1/2 chain 
with an AFM defect of strength $J_D=+3J$ (circles), and with an FM defect of 
strength $J_D=-3J$ (squares). Exact data (full symbols) are compared to LBA 
data (open symbols) and mean-field data (stars).}
\end{figure}

We stress that the LBA calculations for both AFM and FM defects, as well as 
LSA calculations for impurities, can be done analytically, in any 
dimensionality and for any system size, regardless of boundary conditions and 
symmetries. Such local approximations thus provide a very convenient way to 
obtain beyond-mean-field results at the expense of a conventional mean-field 
calculation. When used in this way, local approximations yield robust results 
of moderate precision, even for large and complex systems.  

High precision results, on the other hand, cannot be expected from simple 
local approximations. Improved algorithms for fully numerical diagonalization,
such as that described in Section 3, can be used to make progress from exactly
the opposite starting point: high precision for small systems. 

This work was supported by FAPESP and CNPq.

% The Appendices part is started with the command \appendix;
% appendix sections are then done as normal sections

%\appendix 
%\section{}


\begin{thebibliography}{00}
\bibitem{1} P. Gambardella, Nature Materials {\bf 5}, 431 (2006).
\bibitem{2} J. V. Yakhmi, Physica B {\bf 321}, 204 (2002).
\bibitem{3} W. Kohn, Rev. Mod. Phys. {\bf 71}, (1999) 1253. 
\bibitem{4} V. L. L\'{\i}bero and K. Capelle, Phys. Rev. B {\bf 68}, 
024423 (2003).
\bibitem{5} P. E. G. Assis, V. L. L\'{\i}bero and K. Capelle, 
Phys. Rev. B {\bf 71}, 052402 (2005).
\bibitem{6} K. Capelle and V. L. L\'{\i}bero, Int. J. Quantum Chem. 
{\bf 105}, 679 (2005).
\bibitem{7} V. L. L\'{\i}bero and K. Capelle, Physica B {\bf 384}, 79 (2006).
\bibitem{8} N. A. Lima, M. F. Silva, L. N. Oliveira and K. Capelle, Phys. 
Rev. Lett. {\bf 90}, 146402 (2003).
\bibitem{9} F. C. Alcaraz, M. N. Barber, M. T. Batchelor, R. J. Baxter and 
G. R. W. Quispel, J. Phys. A: Math. Gen. {\bf 20}, 6397 (1987).
\end{thebibliography}
\end{document}